\documentclass[onecolumn,secnumarabic,amssymb, nobibnotes, aps, prb]{revtex4-2}
\usepackage{graphicx}

\begin{document}

\title{Non-magnetic layers with a single Dirac cone at high-symmetry point of the Brillouin zone}%

\author{Vladimir Damljanovi\'c}%
\email[ ]{damlja@ipb.ac.rs}
\affiliation{Institute of Physics Belgrade, University of Belgrade, Pregrevica 118, 11000 Belgrade, Serbia}
\date{February 2020}%
\maketitle

It is well known that a single Dirac cone at high-symmetry point (HSP) of a Brillouin zone, akin to the one in graphenes' band structure, can not appear as the only quasiparticle at the Fermi level in two-dimensional (2D), non-magnetic materials. Here we found two layer groups with time-reversal symmetry, among all possible both without- and with spin-orbit coupling, that host one Dirac cone at HSP and we show which additional dispersions appear: a pair of Dirac lines on opposite BZ edges and a pair of Dirac cones that can be moved but not removed by symmetry preserving perturbations, on the other two BZ edges. We illustrate our theory by a tight-binding band structure and discuss real 2D materials that belong to one of the two symmetry groups. Finally, we single out inconsistencies in the literature showing that it is better to focus scientist attention to a research topics itself, instead to group of authors only (\emph{e.g.} from one university, country etc.) dealing with the same topics. On the other hand, repetitions of other scientist results without due citation, destroys the spirit of science itself.

\section{Introduction}

Two dimensional (2D) materials became focal point of intensive research due to their mechanical and electrical properties suitable both for fundamental research and applications. Materials' 2D Brillouin zone (BZ) allows application of topology in explanation of their physical properties. Electron wave functions near band contacts, like Dirac cones in graphene, gain additional significance when a 2D material is placed in the vertical magnetic field. The absence of back-scattering and the high electron conductivity near materials edges giving rise to Quantum Hall effect are some of the examples \cite{Rev2D, TopMat2}.

The crystal- and time reversal symmetry (TRS) explain both band contacts and dispersion in their vicinity. Group theoretical methodology relies on tables of irreducible (co)representations ((co)reps) of underlying symmetry groups. For dispersions in three-dimensional (3D) crystals, space group representations are firstly used for prediction of bulk chiral fermions \cite{Manjes12}, and finalized in complete list of excitations near high-symmetry point (HSPs) and lines (HSLs) in gray and black-and-white space groups \cite{Enci1, Enci2, Enci3, Enci4}.

Regarding subperiodic groups, computer program POLSym \cite{Pol15} contains (co)reps for all families of line- and layer groups and was used by \cite{Nat1, Nat2, Nat3} to develop spin-group theory for quasi one-dimensional systems. On the other hand, all types of (co)reps of all gray and ordinary layer groups are available recently \cite{dgsite}. Dispersions near all HSPs and HSLs in non-magnetic 2D materials are reported in \cite{Mi23}, which explained published numerical and experimental band structures (\emph{e.g.} in \cite{KoBr20, LiWa19}), and 
can be used as a starting point in designing new 2D materials with physical properties given in advance. Magnetic layer- and rod groups are treated in \cite{Enci2D}, which was published at about the same time as \cite{Mi23}.

Topological laws forbid one Dirac cone, like those in band structure of graphene without spin-orbit coupling (SOC), to be the only feature at a given energy (including the Fermi level) in non-magnetic 2D materials \cite{YK15, WiBr18}. Question arises: in a given symmetry group, if it imposes Dirac cone at one BZ HSP, what other band contacts must appear? Are those contacts essential degeneracies or there are also accidental band contacts (ABCs)? If so, are these ABCs protected by topology or are they truly accidental (can be gaped by symmetry preserving perturbations)? Present paper aims to give answers to these questions.

\section{Method}

List of UNPs and UNLs is given in \cite{Mi23}, for spinless and spinfull electrons in the presence of TRS. The position of UNPs and UNLs is given by the dimensionality of allowed correp of the little group. The matrices of correps can be used to find effective Hamiltonian in the vicinity of UNPs and UNLs, and hence the dispersions (Hamiltonian eigenvalues). We search band structures near unmovable nodal points published in \cite{Mi23}, to identify layer groups having one Dirac cone in the BZ. Next, additional dispersions near UNPs and UNLs that must be present are identified. For investigation of accidental band contacts the non-crossing rule is applied \cite{Laki3, vNW, Hett}. For the tight-binding model, we use the fine structure Hamiltonian $\hat{H}_{\mathrm{fs}}(\mathbf{r})$, obtained from Dirac equation by expansion up to $(1/c)^2$ \cite{Laki4}. More precisely, the following symmetry properties are used \cite{Corn}: $\hat{H}_{\mathrm{fs}}(g^{-1}\mathbf{r})=\hat{u}^{\dagger}(g_0)\hat{H}_{\mathrm{fs}}(\mathbf{r})\hat{u}(g_0)$, where $\hat{u}(g_0)$ represents the rotational part of layer symmetry element $g$ in the spin space and can be found at Bilbao Crystallographic Server \cite{bcsdg}. In addition, TRS implies: $\hat{H}^*_{\mathrm{fs}}(\mathbf{r})=\hat{\sigma}_2\hat{H}_{\mathrm{fs}}(\mathbf{r})\hat{\sigma}_2$, with $\hat{\sigma}_2$ being the Pauli matrix. These symmetry properties of $\hat{H}_{\mathrm{fs}}(\mathbf{r})$ significantly simplify tight-binding Hamiltonian in highly symmetric layers, allowing analytical solution of eigenvalues problem.

\section{Results}

Closer look at dispersions near all HSPs and HSLs \cite{Mi23} for SOC case, gives that centrosymmetric layer double groups $43^D$ ($p\ 2/b\ 2_1/a\ 2/a$) and $45^D$ ($p\ 2_1/b\ 2_1/m\ 2/a$) host one Dirac point at BZ HSP. To these groups corresponds the following space groups: $54^D$ ($P\ 2_1/c\ 2/c\ 2/a$; $y=0$) and $57^D$ ($P\ 2/b\ 2_1/c\ 2_1/m$; $x=0$), respectively (plane parallel to diperiodic one is also given, notation for layer- and space groups is according to \cite{itce} and \cite{itca}, respectively). Besides Dirac point there is also a Dirac line at one edge of the BZ for each of the two groups \cite{Mi23}.

We next investigate if additional band contacts are possible or even unavoidable. We use allowed, extra irreps from Bilbao Crystallographic Server \cite{bcsdg} for corresponding space groups. For group $57^D$, high-symmetry line $H$ has four one dimensional irreps being paired by TRS in the following way: $(\overline{H}_2, \overline{H}_5)$ and $(\overline{H}_4, \overline{H}_3)$. Screw axis of order two is represented by $(-ie^{i\pi u},ie^{i\pi u})$ for both pairs of irreps ($u=1/2$ corresponds to BZ corner). Near end points of line $H$ both pairs are $(-i,i)$ and $(1,-1)$, respectively. However, at the point $Y$ ($u=0$) irreps ($\overline{Y}_3, \overline{Y}_4$) are $(-2i, 2i)$, while those at the BZ corner are $(\overline{T}_3, \overline{T}_4)=(0, 0)$. Since compatibility relations near point $Y$ are not satisfied, two double degenerate zones must touch an odd number of times somewhere on the line $H$ (most probably once). For the remaining parts of the Brillouin zone, accidental band contacts are forbidden by the non-crossing rule. Similar analysis applies to space group $54^D$ with the same result. The main conclusion of this paragraph is that, for layer groups $43^D$ and $45^D$, there are $2(2n-1)$ additional Dirac cones somewhere at the BZ edge (the one that does not host the Dirac line), with $n$ most probably being one. These additional Dirac cones can be moved along the BZ edge, but cannot be gaped by symmetry preserving perturbations.

We illustrate our findings using a tight-binding example on the structure having symmetry $45^D$, shown in the Figure \ref{Figu1} (visualization by VESTA \cite{vesta}). Eight-dimensional tight-binding Hamiltonian in the basis of $s$-orbitals and up and down spinors $\left\{\left|s_1\uparrow\right\rangle, \left|s_2\uparrow\right\rangle, \left|s_3\uparrow\right\rangle, \left|s_4\uparrow\right\rangle, \left|s_1\downarrow\right\rangle, \left|s_2\downarrow\right\rangle, \left|s_3\downarrow\right\rangle, \left|s_4\downarrow\right\rangle\right\}$, is:

\begin{figure}
\includegraphics[width=0.5\textwidth]{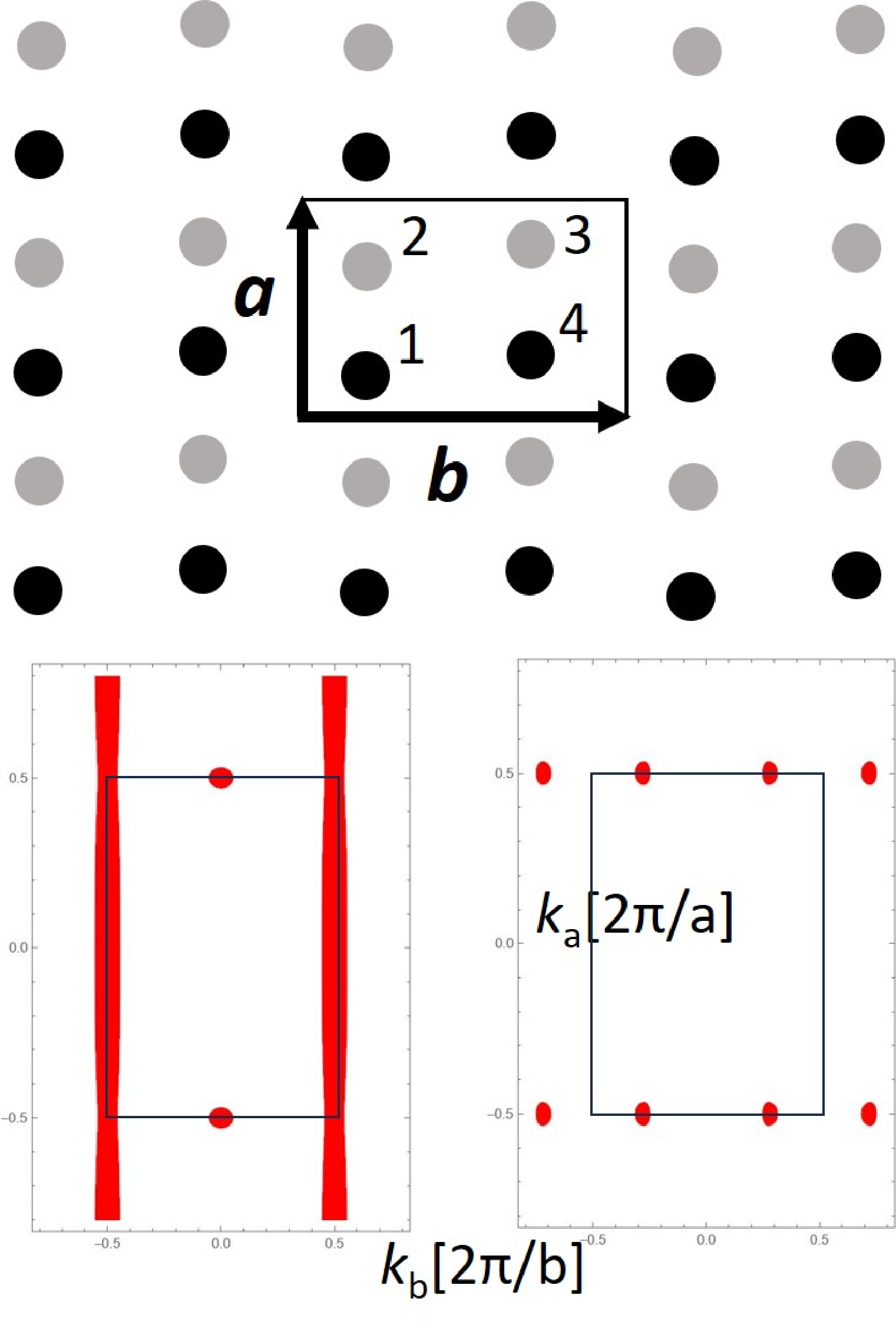}%
\caption{\label{Figu1} Upper panel: crystal structure adopted for the tight-binding model. Black (white) nuclei are above (below) the drawing plane. Lover panels: areas in the reciprocal space (indicated in red color) where $E_2-E_1 < 0.6$ eV (left panel; the same picture is for $E_4-E_3 < 0.6$ eV), \emph{i.e.} $E_3-E_2 < 0.6$ eV (right panel). Black rectangles denote BZ. Parameters $f_0$, $f_1$, $f_2$, $c_1$, $c_2$ are equal to $2.1$, $1.6$, $1.1$, $1.4$, $0.9$ $\mathrm{eV}$, respectively.}
\end{figure}

\label{ham}
\begin{equation}
\hat{H}(\mathbf{k})=\left(
\begin{array}[c]{cc}
	\hat{A}(\mathbf{k}) & \hat{B}(\mathbf{k}) \\
	\hat{B}^{\dagger}(\mathbf{k}) & \hat{A}(\mathbf{k}) 
\end{array}
\right),
\end{equation}
with
\begin{equation}
\hat{A}(\mathbf{k})=\left(
\begin{array}[c]{cccc}
	f_0 & f_1\left(1+e^{-i\mathbf{k}\cdot\mathbf{a}}\right) & 0 & f_2\left(1+e^{-i\mathbf{k}\cdot\mathbf{b}}\right) \\
	f_1\left(1+e^{i\mathbf{k}\cdot\mathbf{a}}\right) & f_0 & f_2\left(1+e^{-i\mathbf{k}\cdot\mathbf{b}}\right) & 0 \\
	0 & f_2\left(1+e^{i\mathbf{k}\cdot\mathbf{b}}\right) & f_0 & f_1\left(1+e^{i\mathbf{k}\cdot\mathbf{a}}\right) \\
	f_2\left(1+e^{i\mathbf{k}\cdot\mathbf{b}}\right) & 0 & f_1\left(1+e^{-i\mathbf{k}\cdot\mathbf{a}}\right) & f_0
\end{array}
\right),
\end{equation}
\begin{equation}
\hat{B}(\mathbf{k})=\left(
\begin{array}[c]{cccc}
	0 & c_1\left(1+e^{-i\mathbf{k}\cdot\mathbf{a}}\right) & 0 & ic_2\left(1-e^{-i\mathbf{k}\cdot\mathbf{b}}\right) \\
	-c_1\left(1+e^{i\mathbf{k}\cdot\mathbf{a}}\right) & 0 & -ic_2\left(1-e^{-i\mathbf{k}\cdot\mathbf{b}}\right) & 0 \\
	0 & ic_2\left(1-e^{i\mathbf{k}\cdot\mathbf{b}}\right) & 0 & c_1\left(1+e^{i\mathbf{k}\cdot\mathbf{a}}\right) \\
	-ic_2\left(1-e^{i\mathbf{k}\cdot\mathbf{b}}\right) & 0 & -c_1\left(1+e^{-i\mathbf{k}\cdot\mathbf{a}}\right) & 0
\end{array}
\right),
\end{equation}
and $f_0$, $f_1$, $f_2$ being real hoping parameters between zeroth, the first, the second neighbors (parallel spins), \emph{i.e.} $c_1$, $c_2$ being real hoping parameters between the first, the second neighbors (antiparallel spins). The electronic band structure, within this model is:
\begin{equation}
\label{disp}
 E_{1,2,3,4}=f_0 \\
\pm2\sqrt{c_1^2\mathrm{cos}^2\left(\frac{\mathbf{k}\cdot\mathbf{a}}{2}\right)+\left[\sqrt{c_2^2\mathrm{sin}^2\left(\frac{\mathbf{k}\cdot\mathbf{b}}{2}\right)+f_1^2\mathrm{cos}^2\left(\frac{\mathbf{k}\cdot\mathbf{a}}{2}\right)}\pm\left|f_2\mathrm{cos}\left(\frac{\mathbf{k}\cdot\mathbf{b}}{2}\right)\right|\right]^2}, \nonumber
\end{equation}
with each of the four bands from (\ref{disp}) being doubly degenerate, as expected from Kramer’s theorem for centrosymmetric systems with SOC. Four doubly degenerate bands are obtained by four possible combinations of $+$ and $-$ in (\ref{disp}). Bands $E_1$, $E_2$, $E_3$, $E_4$ are ordered according to non-decreasing energy for each $\mathbf{k}$ in the following way: $(-,+)\leq(-,-)\leq(+,-)\leq(+,+)$. The four-fold degeneracy is obtained for:
\begin{equation}
c_2^2\mathrm{sin}^2(\mathbf{k}\cdot\mathbf{b})+4f_1^2\mathrm{cos}^2\left(\frac{\mathbf{k}\cdot\mathbf{a}}{2}\right)\mathrm{cos}^2\left(\frac{\mathbf{k}\cdot\mathbf{b}}{2}\right)=0,
\end{equation}
which is valid at $\mathbf{k}\cdot\mathbf{b}=\pm\pi$ (unmovable Dirac lines).

Along lines $\mathbf{k}\cdot\mathbf{a}=\pm\pi$ dispersions (\ref{disp}) become:
\begin{equation}
E_{1,2,3,4}=f_0\pm2\left|\left|f_2\mathrm{cos}\left(\frac{\mathbf{k}\cdot\mathbf{b}}{2}\right)\right|\pm\left|c_2\mathrm{sin}\left(\frac{\mathbf{k}\cdot\mathbf{b}}{2}\right)\right|\right|,
\end{equation}
which gives $\mathbf{k}\cdot\mathbf{b}=0$ (unmovable Dirac points) and $\mathbf{k}\cdot\mathbf{b}=\pm2\mathrm{arctg}(f_2/c_2)$, as positions of movable Dirac points from touching of two inner, doubly degenerate bands.

To investigate dispersions near touching points closer, we write $\mathbf{k}=\mathbf{k}_0+\mathbf{q}$ and expand (\ref{disp}) for small $|\mathbf{q}|$. Dispersions become:
\begin{equation}
E_{1,2,3,4}(\mathbf{q})\approx f_0\pm2|f_2|\pm\sqrt{f_1^2(\mathbf{q}\cdot\mathbf{a})^2+c_2^2(\mathbf{q}\cdot\mathbf{b})^2},
\end{equation}
near unmovable Dirac points, \emph{i.e.}:
\begin{equation}
E_{3,2}(\mathbf{q})\approx f_0\pm\sqrt{c_1^2(\mathbf{q}\cdot\mathbf{a})^2+(c_2^2+f_2^2)(\mathbf{q}\cdot\mathbf{b})^2},
\end{equation}
near movable Dirac points. The absence of other contacts between bands across the reciprocal space is illustrated in Figure \ref{Figu1}, which is in accordance with the non-crossing rule. It also follows from Figure \ref{Figu1} that all four, doubly degenerate bands are tangled together so that an insulator with such a band structure must have total number of electrons per primitive cell divisible by eight (unless it is strongly correlated). This is in accordance with filling factor for layer group $45^D$ \cite{ElFil3}, \emph{i.e.} it's corresponding space group \cite{ElFil1, ElFil2}. The same filling factor is reported for layer group $43^D$ \cite{ElFil3, ElFil1, ElFil2}.

\section{Discussion and conclusions}

We discuss our results in relations to the existing literature and in a broader context of this research area. The positions of movable nodal points and lines published for 3D hexagonal \cite{AccBCh}, orthorhombic \cite{AccBCo} and tetragonal \cite{AccBCt} space groups could be used for layer groups too, via corresponding space groups. Movable band contacts for groups $43^D$ and $45^D$ found here are reported for their 3D analogs $54^D$ and $57^D$ \cite{AccBCo}. When the third dimension to \emph{k}-space is added, the band contacts remain point-like for $54^D$, but become part of a movable nodal line for $57^D$ \cite{AccBCo}.

Using tight-binding model, authors of \cite{YK15} classify band touchings in non-magnetic materials into four possible cases. All four cases are obtained by breaking symmetry elements in layer group $64^D$ \cite{YK15}. Groups $43^D$ and $45^D$ arise by further breaking symmetry of case I (two symmetry equivalent Dirac points), from \cite{YK15}. Group-subgroup chain is $43^D < 48^D < 64^D$ and $45^D < 48^D < 64^D$, so these groups are not maximal subgroups of $64^D$ \cite{bcssubg, bcssubg2}. This case is not included into classification of \cite{YK15}. The same holds for centrosymmetric, non-symmorphic groups $40^D$, $44^D$ and $63^D$, hosting Dirac lines instead of cones.

For experimental realization of materials belonging to layer groups given in advance, one can search database of 2D materials \cite{DBase18, DBase21} and layered 3D materials which were synthesized in the past \cite{Exp}. Material $\mathrm{KAsO_2}$ belongs to layer group $45$, with synthesis of it's 3D layered analog described before \cite{ExpPrim}. The 2D version is non-magnetic insulator with a band gap of around $4$eV \cite{DBase18}. Further investigations would be necessary to show if it would be possible to shift it's Fermi level by \emph{e.g.} doping. On the other hand, band connectivity of groups $43^D$ and $45^D$ (and their common subgroup $33^D$), which is the highest among all layer groups, forbids their realization by small deformations of a material belonging to any of their supergroups. Deformations, however small they may be, are supposed to close a finite gap at some BZ point, which we think is unlikely.

Groups $40^D$, $43^D$ and $45^D$ are reported \cite{KiGr23} to host semi-Dirac fermions (SDF), which disperse linearly in one direction in \emph{k}-space and quadratically in the orthogonal direction. Indeed, tight binding dispersions (\ref{disp}) become $E_{1,2,3,4}\approx h_{\pm}+h_1^{\pm}(\mathbf{q}\cdot\mathbf{a})^2\pm h_2|(\mathbf{q}\cdot\mathbf{b})|$, with $h_{\pm}=f_0\pm2\sqrt{c_1^2+c_2^2+f_1^2}$, $h_1^{\pm}=\mp(c_1^2+f_1^2)/(4\sqrt{c_1^2+c_2^2+f_1^2})$, $h_2=|f_2|\sqrt{c_2^2+f_1^2}/\sqrt{c_1^2+c_2^2+f_1^2}$ (near points $(0,\pm\pi/b)$), \emph{i.e.} $h_{\pm}=f_0\pm2|c_2|$, $h_1^{\pm}=\pm(c_1^2+f_1^2)/(4|c_2|)$, $h_2=|f_2|$ (near points $(\pm\pi/a,\pm\pi/b)$); only upper (only lower) signs are combined in $h_{\pm}$, $h_1^{\pm}$. However, list of groups and BZ points hosting SDF in \cite{KiGr23} is incomplete. As shown in \cite{Mi23}, SDFs appear also at the midpoints of BZ borders of groups $44^D$ and $63^D$ and at the BZ corners of group $43^D$. Also, the symmetry adapted low energy Hamiltonian derived in \cite{KiGr23}, gives point-like, instead of line-like degeneracies required by symmetry. The term SDF is used for line-nodes extrema (like found in \cite{JaSe17}), which differs from semi-Dirac dispersion around point-like degeneracy as introduced in \cite{SeDi09a, SeDi09b}.

When SOC is neglected, the hoping parameters between states with opposite spins vanish, while those with the parallel spins slightly change. The dispersions (\ref{disp}) become $\widetilde{E}_{1,2,3,4}=\widetilde{f}_0\pm2\left||\widetilde{f}_1\mathrm{cos}(\mathbf{k}\cdot\mathbf{a}/2)|\pm|\widetilde{f}_2\mathrm{cos}(\mathbf{k}\cdot\mathbf{b}/2)|\right|$, which give fortune teller states (FT) near BZ corners $\widetilde{E}_{1,2,3,4}=\widetilde{f}_0\pm\left||\widetilde{f}_1(\mathbf{q}\cdot\mathbf{a})|\pm|\widetilde{f}_2(\mathbf{q}\cdot\mathbf{b})|\right|$, as predicted before for layer (single) groups $33$, $43$ and $45$ \cite{Ja17}. Non-zero density of states (DOS) of $2/\left(\pi\mathrm{max}\left\{b|\widetilde{f}_1|,a|\widetilde{f}_2|\right\}\right)$ near bands contact, distinguish FT dispersion from other linear dispersions. The formula for DOS presented here does not include doubly degeneracy of bands and corrects inessential numerical error in \cite{Ja17}. Eight-fold spinfull degeneracy for electrons (\emph{i.e.} four-fold degeneracy for \emph{e.g.} phonons) at BZ corners, together with FT dispersion, is guaranteed by mere belonging of a 2D material to one of these three groups \cite{Ja17}. Later it was pointed out that in some layer single groups, four-fold degeneracy (eight-fold, for electrons with spin) might appear also on some high-symmetry lines \cite{OsKin, OsKinE} by truly accidental band contacts (TABC). Although being an interesting idea, such TABC can be destroyed by symmetry preserving perturbations and are assumed to appear rarely in real materials in \cite{Ja17}. In other words, even if present by chance, TABC could be gaped by change of temperature, pressure, humidity \emph{etc}, which could cause a potential device designed by use of TABC to malfunction. The absence of TABC on BZ edges in our tight-binding model without SOC, in \emph{ab initio} band structure of a silicone monolayer in \cite{Ja17} and measured by ARPES \cite{Zdyb}, are another confirmations of our statements. Note that part of Table I of \cite{DirPh22} (apart from materials mentioned in it) that deals with Dirac phonons at HSPs, repeats group-theoretical results published five years before in Table I of \cite{Ja17} for unmovable, four-fold (spinless) degenerate band contacts. The other part (Dirac phonons on HSLs \cite{DirPh22}), contains seven groups that might host TABC, six of which being reported and one originally omitted in \cite{OsKin, OsKinE}.

Some errors found in the literature could have been avoided, had the authors red (and consequently, cited) previous publications on the same or similar topic. Reference \cite{KinGr23} study phonons in $\mathrm{PtI_4}$ and $\mathrm{LiBiO_2}$ monolayers belonging to layer groups $33$ and $45$, respectively. The $\mathbf{k}\cdot\mathbf{p}$ Hamiltonian derived in \cite{KinGr23} around the BZ corner of group $45$, does not reduce to zero at BZ corner and has four real parameters, instead of three required by symmetry \cite{Ja17}. The number of real independent parameters in a low-energy Hamiltonian derived from symmetry is invariant, unlike it's form which depends on the basis, and could be used as a crosscheck. Group theory was used for finding which gray layer groups host Dirac cones both without and with SOC \cite{KinGr20}. Group $73$ is missing in results of \cite{KinGr20}, as can be seen by comparing cases with and without SOC in \cite{Mi23}. Gray layer groups without SOC hosting Dirac cones listed much earlier in \cite{Ja16, Ja16Ad}, might be a useful starting point of analysis performed in \cite{KinGr20}. Here we add that eleven groups being main group-theoretical result summarized in Table I of \cite{KinNecit23}, are the exact eleven groups published seven years before in \cite{Ja16, Ja16Ad}. Group theoretical treatment of Dirac points of electrons without SOC (treated in \cite{Ja16, Ja16Ad}) and linear Weyl points of phonons (treated in \cite{KinNecit23}) is identical. Similarly, the group-theoretical results of \cite{Mi20}, reporting electronic dispersions near four-fold degeneracies in non-centrosymmetric materials with SOC, are repeated in \cite{KinNecit22}. The $\mathbf{k}\cdot\mathbf{p}$ Hamiltonian near $\overline{M}$ point of wallpaper group $\mathrm{p4g}$ (which is also layer group $56$) has two real parameters and lead to doubly-degenerate bands as if the group was centrosymmetric \cite{KinNec22}. This is not in accordance with earlier publication \cite{Mi20}, where the Hamiltonian near BZ corners of layer group $56$ has three parameters and gives doubly degeneracy only along HSLs, as required by symmetry. These examples show that it is not good to limit attention to a group of authors only (\emph{e.g.} from one university, country etc.), instead to a topic of research itself. Also, repetition of previous results of other authors without due citation, destroys the spirit of science itself.

Finally, we mention that our group theoretical treatment of dispersions holds for fixed nuclei and should not be transferred directly to dynamical Hamiltonians, such as the electron-phonon Hamiltonian relevant for explanation of conventional superconductivity. For dynamical problems, symmetry analysis of phonons (like the one for graphene \cite{Dres}), could be useful. Also, a single Dirac cone at HSP as the only quasiparticle at the Fermi level, appears in some black-and-white layer groups describing antiferromagnetic ordering \cite{SinDP17}, where time reversal alone is not a symmetry.

\subsection{Acknowledgments}
Author acknowledges funding by the Ministry of Science, Technological Development and Innovation of the Republic of Serbia provided by the Institute of Physics Belgrade.

%


\end{document}